\documentclass[12pt]{iopart} 
\usepackage{iopams}  
\usepackage{graphicx}
\usepackage{dcolumn}
\usepackage{bm}
\usepackage{hyperref}

\begin{document}
\title{Transport dynamics of ultracold atoms in a triple-well transistor-like potential}

\author{Seth C. Caliga$^1$, Cameron J. E. Straatsma$^2$, Dana Z. Anderson$^1$}
\address{$^1$ JILA and Department of Physics, University of Colorado and National Institute for Standards and Technology, Boulder CO 80309-0440, USA}
\address{$^2$ JILA and Department of Electrical, Computer, and Energy Engineering, University of Colorado, Boulder, CO 80309-0440, USA}

\begin{abstract}
The transport of atoms is experimentally studied in a transistor-like triple-well potential consisting of a narrow gate well surrounded by source and drain wells. Atoms are initially loaded into the source well with pre-determined temperature and chemical potential. Energetic atoms flow from the source, across the gate, and into the drain where they are removed using a resonant light beam. The manifestation of atom-atom interactions and dissipation is evidenced by a rapid population growth in the initially vacant gate well. The transport dynamics are shown to depend strongly on a feedback parameter determined by the relative heights of the two barriers forming the gate region. For a range of feedback parameter values, experiments establish that the gate atoms develop a larger chemical potential and lower temperature than those in the source.
\end{abstract}
\maketitle

\section{\label{sec:Intro}Introduction\protect\\}
Atomtronic devices utilize tailored potential energy landscapes to control the flow of neutral atoms in a manner that mimics the flow of current through analogous electronic devices~\cite{Seaman:2007kx,Stickney:2007ix,Pepino:2009jb,Pepino:2010,Moulder:2012,Zozulya:2013te,Caliga:2013mw,Ryu:2013,Eckel:2014}.  Atom flow can be driven by gradients in temperature and chemical potential in much the same way that electron flow can be driven in an electronic circuit. Interatomic interactions substitute for Coulomb repulsion, giving rise to a contribution to the chemical potential that constitutes the dual of electric voltage.  The role of thermodynamic gradients in transport dynamics has been studied in ultracold atom systems, at both zero and nonzero temperature~\cite{Shin:2004,Brantut:2013he,Lee:2013,Hazlett:2013,Labouvie:2015nd}.  In such systems, atom currents driven by chemical potential and temperature gradients attempt to bring the system into thermodynamic equilibrium.  Whereas classical electronic circuits are well coupled to a thermal reservoir, typically keeping temperatures close to that of the ambient environment, atomtronic circuits are necessarily isolated from their surroundings; thus, thermal effects play a determining role in circuit dynamics. For this reason, understanding transport processes within finite temperature atomic systems having synthesized potential energy landscapes is relevant to the development of atomtronic devices and circuits.  Given the ubiquity of the transistor in classical electronic signal processing, the realization of an atomtronic analogue of the transistor for use in quantum information and signal processing is of particular interest.  

This paper presents an experimental investigation of the flow of atoms within a triple-well potential, shown in Fig.~\ref{fig:potential}, which is similar to the potential in previous studies of atomtronic transistor behavior~\cite{Stickney:2007ix,Caliga:2013mw}. The two barriers that separate the wells are analogous to the two junctions of the canonical semiconductor transistor. In the experiments, atoms are prepared in the left-hand well while the other two are initially empty. The growth of population in the central well is studied in order to elucidate the role of dissipative processes associated with thermalization. Atoms that traverse both barriers are removed by resonant light. This form of localized dissipation enables the study of continuous non-equilibrium dynamics~\cite{Labouvie:2015ne}. By varying the relative height of the two barriers we demonstrate control over the steady-state chemical potential and temperature of the central well. Within an electronic device, such control is equivalent to biasing node voltages and setting physical device parameters, which is necessary for establishing proper device operation. For example, forward and reverse biases are applied to a semiconductor transistor to achieve current gain.  Similarly, the operation of an atomtronic transistor depends on differences in chemical potential and temperature throughout the system.  Transistor action in a triple-well, atomtronic transistor system, operated around fixed forward or reverse bias points, is reported on elsewhere~\cite{Caliga:2015}.  In that related work, regions of either positive or negative differential current gain are shown to occur when the central well is forward or reverse biased, respectively, and an external current is injected into the central well.  The transport dynamics presented in this paper demonstrate the ability to forward or reverse bias the central well of the triple-well system, and that the bias is controlled by the relative height of the two barriers.  

\section{\label{sec:Intro}Experiment\protect\\}
\begin{figure}
\begin{center}
\includegraphics[width=4.5in]{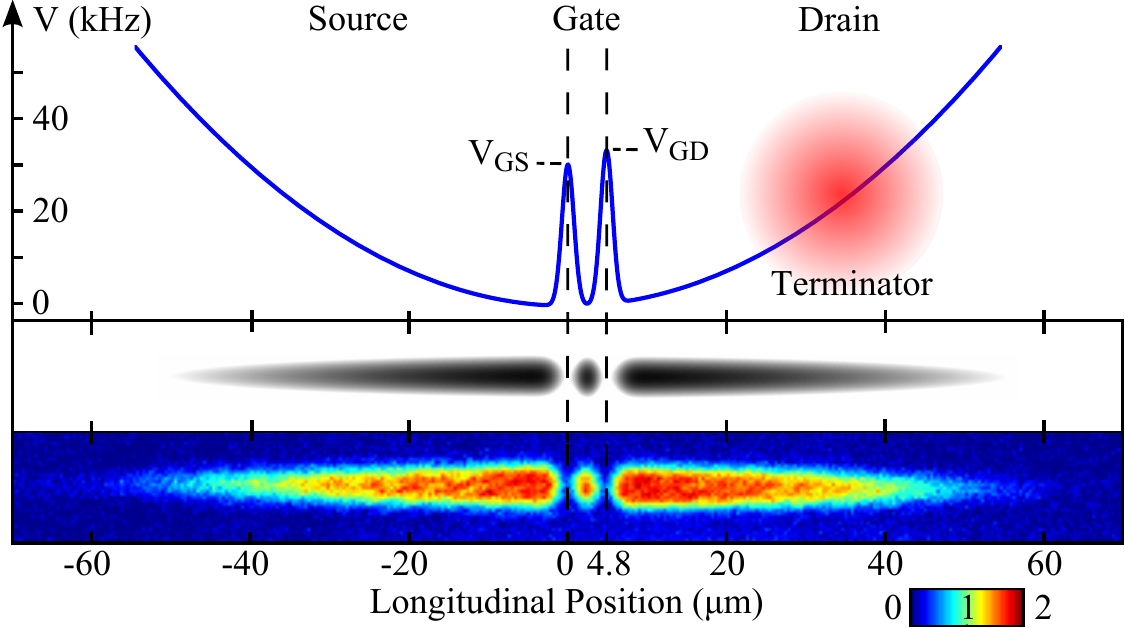}
\caption{\label{fig:potential}A hybrid magnetic and optical potential provides confinement for the atomic ensemble. The upper panel shows the loose longitudinal magnetic confinement along with the $2.1~\mu\mathrm{m}$ full-width at 1/e, blue-detuned optical barriers separated by $4.8~\mu\mathrm{m}$. The barriers are made to be $16~\mu\mathrm{m}$ long by raster scanning the beam along the radial direction of the magnetic trap. $\mathrm{V}_{\mathrm{GS}}$ and $\mathrm{V}_{\mathrm{GD}}$ represent the height of the two barriers separating the source and gate wells and the gate and drain wells, respectively. The terminator beam, illustrated in the drain well, out-couples atoms from the trap by optically pumping them into an untrapped $m_F$ state. The middle panel is a calculated 2D potential energy plot, and the bottom panel shows a false color in-trap absorption image of atoms occupying all three wells, with the terminator beam absent. An optical density scale is shown below the horizontal axis.}
\end{center}
\end{figure}

Borrowing the nomenclature of field-effect transistors we label the three regions of our triple-well potential as the ``source," ``gate," and ``drain"~\cite{Nom:Trans}. These regions are indicated within the longitudinal profile of the trapping potential, shown in the top of Fig.~\ref{fig:potential}.
Throughout this paper energies other than temperature are reported in units of Hz. The bottom panel of Fig.~\ref{fig:potential} provides a false color in-trap absorption image of approximately $4.5\times10^4~^{87}\mathrm{Rb}$ atoms that have been loaded into the trap and subsequently cooled to a temperature, $T$, near the critical temperature, $T_\mathrm{c} \approx 1~\mu\mathrm{K}$, by forced radio-frequency (RF) evaporation. The image reveals the three distinct regions of the triple-well potential that are created using a combination of magnetic and optical trapping forces.  An atom chip in conjunction with bias fields generated by external coils produces a cigar-shaped magnetic trap located $\sim150~\mu$m below the surface of the atom chip.  Optical access to the trap is enabled by a transparent, coplanar region of glass embedded in the silicon chip substrate~\cite{Salim:2013gb}.  A pair of blue-detuned ($\lambda = 760$ nm) optical barriers are projected through the transparent chip window, sectioning the magnetic potential longitudinally into three wells.

\begin{figure}
\begin{center}
\includegraphics[width=5.5in]{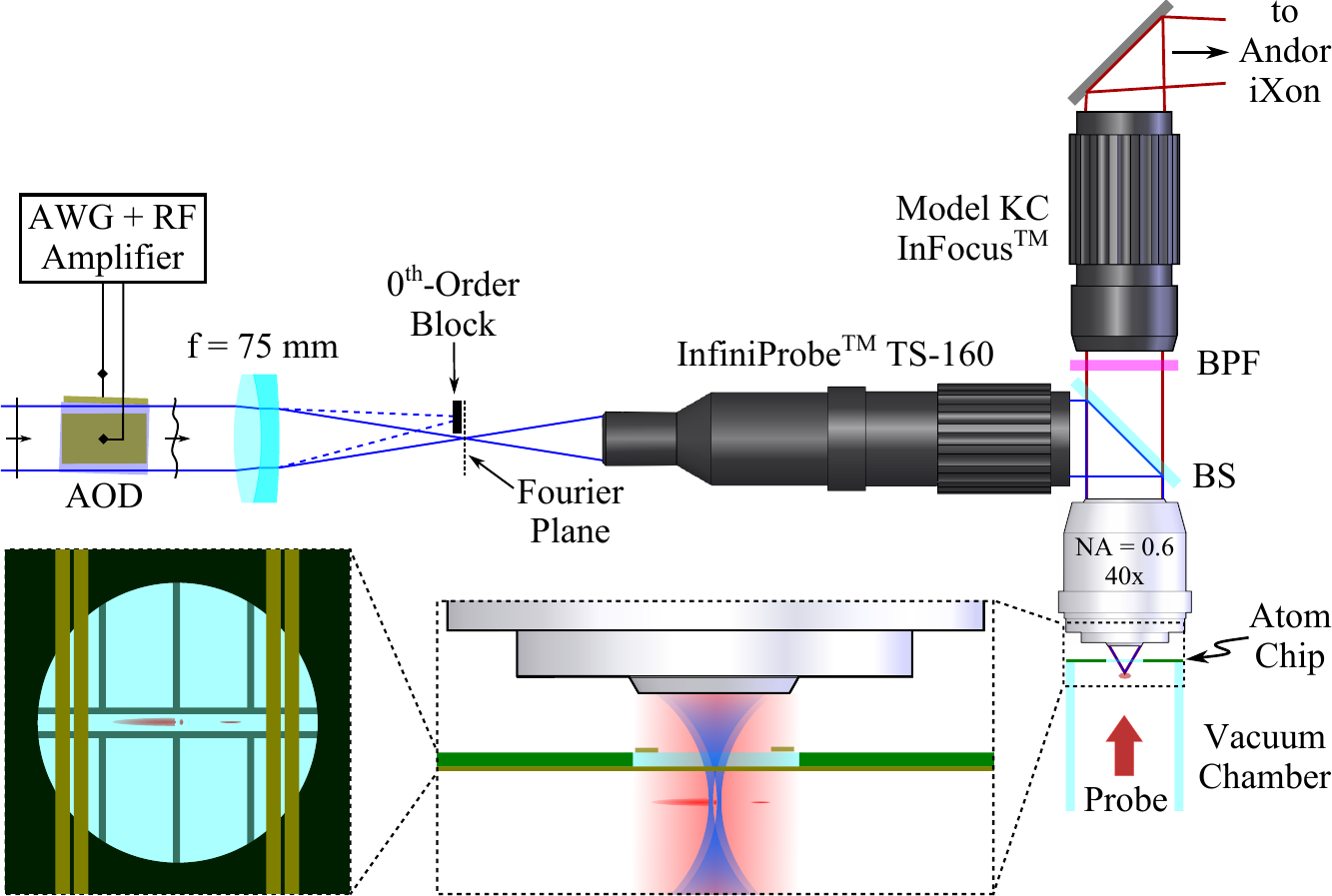}
\caption{\label{fig:optical_system} Schematic of the microscope system used for studying transport processes in a triple-well potential. Detailed operation is described in the text. Red and blue beam paths indicate imaging and barrier projection paths, respectively. An expanded view of through-chip absorption imaging and barrier projection is shown.  Trapped atoms, not drawn to scale, are shown relative to the 3 mm diameter window and conductor layout of the atom chip. BS: beam splitter, BPF: band pass filter ($\lambda_\mathrm{c}=780~$nm, $\Delta\lambda=1~$nm)}
\end{center}
\end{figure}

A schematic of the microscope system used for our experiments is shown in Fig.~\ref{fig:optical_system}.  
The system is constructed from commercially available components and achieves high numerical aperture performance without the need for custom in-vacuo optics. The use of an infinity-corrected objective lens and a beam splitter allow for simultaneous in-trap absorption imaging and the projection of the optical barriers. Atoms are imaged in-trap using resonant probe light that illuminates the atoms from below and is subsequently collected by a Zeiss LD Plan-Neofluar objective ($\mathrm{NA} = 0.6$ and $40\times$ magnification) located outside of the vacuum cell, approximately $3~\mathrm{mm}$ above the atom chip. Note that this objective was chosen as it contains a coverslip correction collar, which is used to remove aberrations due to the $420~\mu$m thick atom chip window. The probe light passes through one port of a beam splitter followed by an Infinity Photo-Optical Model KC InFocus\textsuperscript{TM} lens system, which is used to form an image of the atoms on an Andor iXon electron multiplying charge-coupled device. With the setup shown in Fig.~\ref{fig:optical_system}, we achieve a diffraction limited spot size of $1.6~\mu\mathrm{m}$ (Airy disk diameter) and an object-space pixel size of $0.4~\mu\mathrm{m}$. The optical barriers are created using a 2D acousto-optic deflector (AOD) driven by a two-channel arbitrary waveform generator (AWG), which allows the two barrier positions and heights to be adjusted independently. A doublet lens Fourier transforms the output of the AOD and forms the desired optical potential pattern at the front focal plane of an Infinity Photo-Optical InfiniProbe\textsuperscript{TM} TS-160 lens system. The InfiniProbe\textsuperscript{TM} images the optical pattern of the barriers through the other port of the beam splitter where the barriers are projected onto the atoms using the same objective lens used for absorption imaging. 

\begin{figure}
\begin{center}
\includegraphics[width=4.5in]{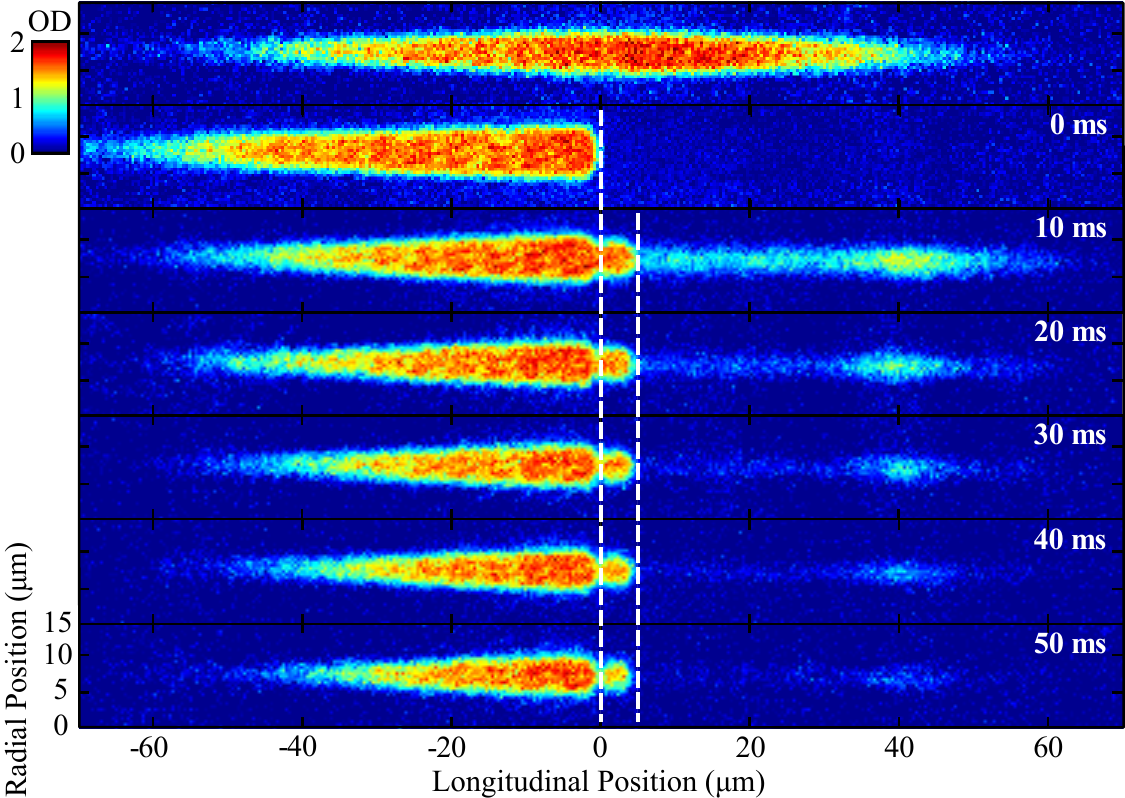}
\caption{\label{fig:intrap_time} In-trap absorption images showing state preparation and temporal evolution of the system from $\Delta t = 0~\mathrm{ms}$ to $\Delta t = 50~\mathrm{ms}$ in $10~\mathrm{ms}$ intervals. White dashed lines indicate the barrier positions throughout the experimental sequence. Each image is an average of 5 separate realizations of the experiment. Note the rapid rise of atom density in the gate well after only $10~\mathrm{ms}$.  Short length-scale structure in saturated regions is due to minor interference effects in the imaging system.}
\end{center}
\end{figure}

The transport dynamics experiments described in this paper are initialized by creating a finite temperature Bose-Einstein condensate (BEC) in the bare magnetic trap using forced RF evaporation.  Next, all of the atoms are swept into the source well using the optical barriers by modulating the waveform that controls the longitudinal barrier positions (see Fig.~\ref{fig:intrap_time}). During this preparation stage the barriers are kept sufficiently high such that the gate and drain wells are initially empty. The longitudinal frequency of the bare magnetic trap (i.e. with no barriers present) is $\nu_x = 67~\mathrm{Hz}$, and with the barriers present we approximate the source well as a half harmonic well. The half harmonic well approximation is used to determine the chemical potential in the Thomas-Fermi (TF) limit. This approximation for the source well leads to an error of $<5\%$ in the chemical potential as confirmed with 3D simulations of the Gross-Pitaevskii equation for a BEC in the ground state of the actual potential. For a barrier separation of $4.8~\mu\mathrm{m}$ the longitudinal trap frequency of the gate is $\nu_{x,G} \approx 850~\mathrm{Hz}$.  In conjunction with the Gaussian profile of the barriers, this spacing minimizes the anharmonicity of the central well. Note that the degree of overlap between the two optical barriers contributes to an offset in the potential energy of the gate well. The shift in the minimum energy of each well due to the Gaussian shape of the barriers is denoted by $V_{0,i}$. For the data we present, the barrier separation of $4.8~\mu\mathrm{m}$ results in $V_{0,S}\approx V_{0,G}\approx V_{0,D}\approx 0$. The radial frequency of all three wells is that of the bare magnetic trap, $\nu_\perp = 1500~\mathrm{Hz}$.  

We begin each experimental realization by initializing the source well with a total atom number $N_\mathrm{s} = 20.0(2)\times10^3$ atoms at a temperature $T_\mathrm{s} = 720(25)$~nK, corresponding to a TF chemical potential of $\mu_s = 3.0(2)$~kHz. At time $t = 0$ the barrier heights are set to the desired $V_\mathrm{GS}$ and $V_\mathrm{GD}$ and the system is allowed to evolve for a time $\Delta t$. During this evolution time the drain well is illuminated with laser light tuned to atomic resonance (labeled ``terminator" in Fig.~\ref{fig:potential}) such that any atoms reaching the drain are removed from the trap.  Thus, the terminator beam prevents the system from equilibrating and ensures that no current flows from drain to gate, which would alter the gate population.  The terminator beam serves an analogous purpose to termination in RF electronics, where proper termination impedance matches the device output to the load circuit in order to eliminate signal reflection and subsequent interference.  The frequency and polarization of the terminator beam are set to pump atoms from the magnetically trapped $F = 2,~m_\mathrm{F}=2$ state to any of the untrapped $F = 2,~m_\mathrm{F}\leq0$ states. The terminator beam is projected along the same beam path as the optical barriers and has a typical power of $\sim 10^{-12}$ W. It is displaced by $32~\mu\mathrm{m}$ into the drain well and has a full width at $1/\mathrm{e}$ of $16~\mu\mathrm{m}$.  Measurements on the state of the system are performed using standard absorption imaging techniques, either in time-of-flight (TOF) or in-trap. Data taken in TOF provides quantitative information about the temperature and chemical potential of the atoms in the source, while in-trap data provides information on the occupancy of each well. For in-trap imaging, the terminator beam is extinguished by a time $\sim 1/4\nu_x\approx 4~$ms prior to acquiring an image, such that we obtain a snapshot of the atom current flowing into the drain well.

Figure~\ref{fig:intrap_time} shows a series of in-trap absorption images including the initial state preparation and evolution for $\Delta t = 0 - 50~\mathrm{ms}$ during which $\mathrm{V}_{\mathrm{GS}} = 30~\mathrm{kHz}$ and $\mathrm{V}_{\mathrm{GD}} = 33~\mathrm{kHz}$.  
This series of images was recorded using a weak probe beam with intensity $I_\mathrm{p} \lesssim I_\mathrm{sat}$, where $I_{sat}$ is the saturation intensity of the atomic transition,  in order to observe atoms that have reached the drain. As can be seen in the third frame of the series, atoms become trapped in the gate well by $\Delta t = 10$ ms. If atoms were not trapped in the gate, but rather only traversing it, one would expect there to be at most twice the density seen in the drain well, just to the right of the gate-drain barrier.  We note that at early evolution times atoms that enter the gate well experience a region of substantial population inversion since energy states lying below the barriers are initially unpopulated. While the route to steady-state is an analytically difficult problem, the role of interatomic interactions and manifestation of dissipation are made evident by the population growth in the gate well.

The time series shown in Fig.~\ref{fig:intrap_time} also reveals the accumulation of atoms at the far end of the drain well.  By extinguishing the terminator beam early, atom current into the drain is allowed to propagate to the classical turning point of the trap dictated by the energy of emitted atoms.  The longitudinal position and spatial distribution yield information regarding the height of the gate-drain barrier and the momentum spread, respectively, of atoms entering the drain well. The longitudinal momentum spread of the flux of atoms into the drain is determined by fitting a Gaussian to the longitudinal spatial distribution that peaks at $\sim 40~\mu$m, which corresponds to a spread in kinetic energy, $\Delta E$.   We find that $\Delta E \approx kT_\mathrm{s}$, where $k$ is the Boltzmann constant, indicating that the current into the drain has an energy spread on the order of the temperature of the ensemble in the well from which they originate.

\begin{figure}
\begin{center}
\includegraphics[width=5.5in]{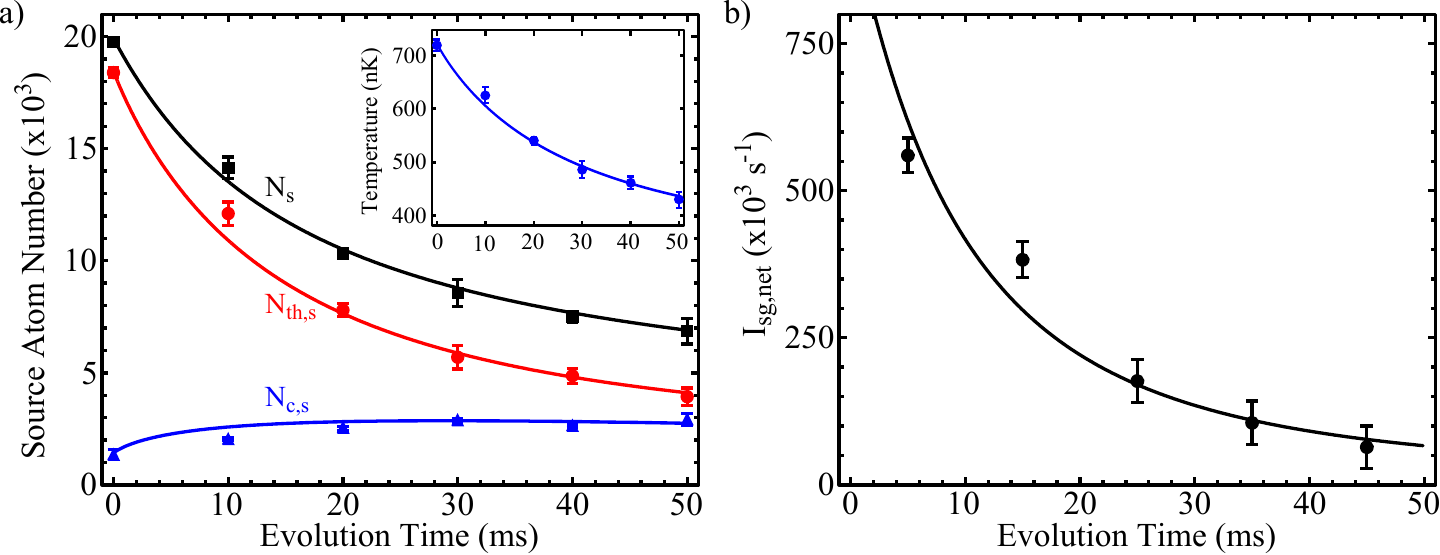}
\caption{\label{fig:atom_number} a) Source well total (black square), thermal (red circle), and condensed (blue triangle) atom numbers measured in TOF corresponding to the in-trap images in Fig.~\ref{fig:intrap_time}. The inset shows the temperature for the same data. Error bars indicate the standard error of the mean of five experimental realizations at each evolution time. b) Net source-gate atom current determined from the first time derivative of the source well total atom number. Error bars indicate the error propagated from data in (a). Solid lines in both (a) and (b) show the corresponding values calculated numerically from Eqs.~(\ref{eqn:current}) and~(\ref{eqn:energy}).}
\end{center}
\end{figure}
We extract quantitative data on the thermodynamic evolution of the source ensemble using TOF absorption imaging.  Figure~\ref{fig:atom_number}(a) shows the total, $N_\mathrm{s}$, thermal, $N_\mathrm{th,s}$, and condensed, $N_\mathrm{c,s}$, atom numbers determined from a bimodal fit to the momentum distribution of the cloud in free expansion.  The decay of the total atom number in the source well reflects the flux of atoms emitted into the gate, and can be used to quantify the source-gate atom current.   The growth of the source well condensate atom number indicates that the atom current is comprised of thermal atoms. The increasing condensate fraction, which approaches $N_\mathrm{c,s}/N_\mathrm{s} \approx 0.5$ by $\Delta t = 50$~ms, is accompanied by a decrease in the temperature of the source ensemble as shown in the inset of Fig.~\ref{fig:atom_number}(a). Removal of higher energy thermal atoms and subsequent cooling of those that remain is akin to evaporative cooling.  However, rather than removing the atoms from the system completely, as is the case in evaporative cooling, the triple-well potential directs the atom current first into the gate well, and subsequently into the drain. Thus, as current flows the cooling that occurs in the source well is indicative of the power delivered to the drain.  

\begin{figure}
\begin{center}
\includegraphics[width=5.5in]{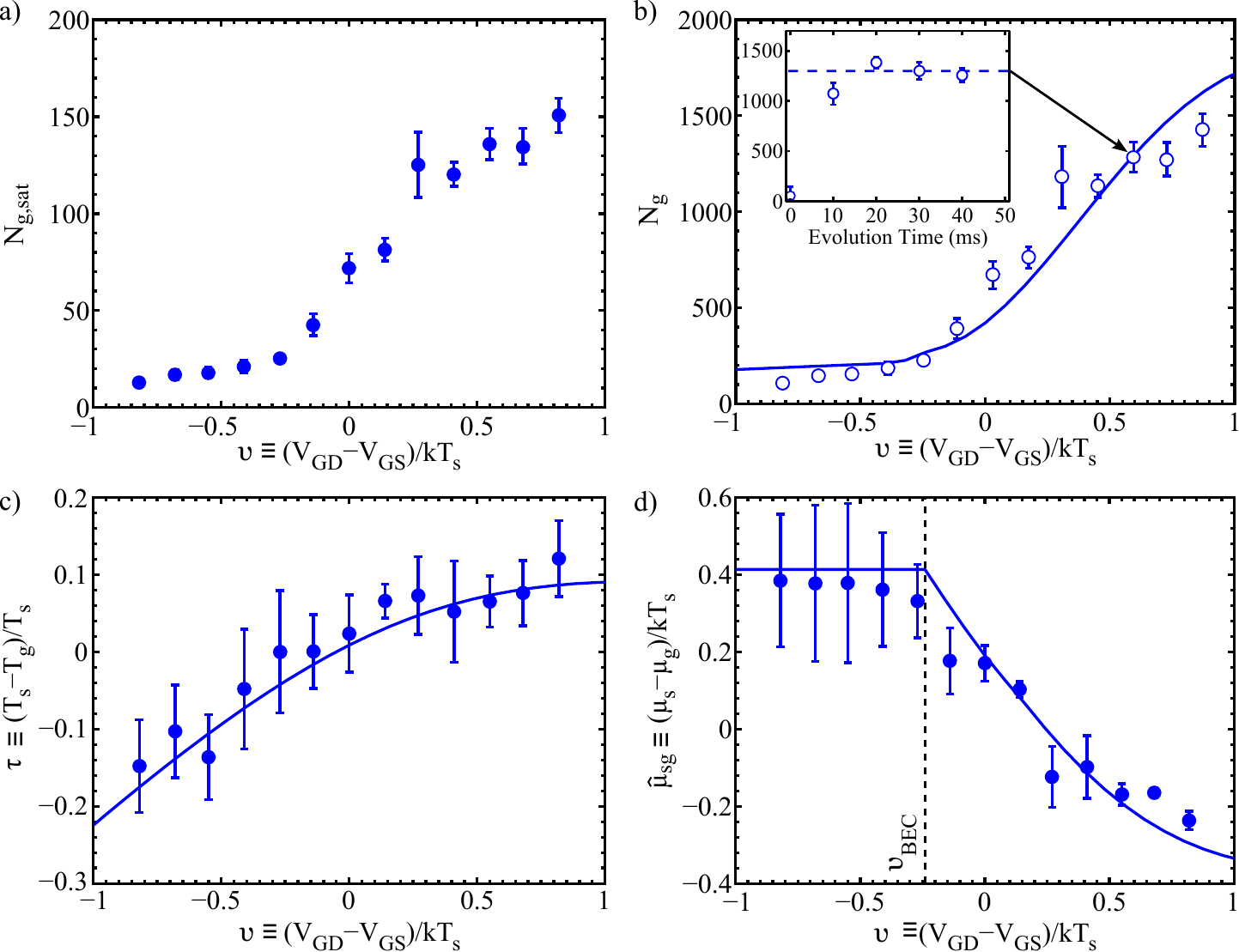}
\caption{\label{fig:gate_turnon} Gate filling behavior observed in-trap: a) Total gate well population, $N_\mathrm{g,sat}$, measured using strongly saturated absorption imaging as a function of the feedback parameter.  Error bars indicate the standard error of the mean of five experimental realizations at each feedback parameter. b) Numerically calculated steady-state gate well population, $N_\mathrm{g}$, (solid line) and data from (a) scaled by a dimensionless factor, $C = 9.3(1)$ (open circles). The inset shows the time evolution of the gate population (open circles) relative to the steady-state population (dashed line). c) Temperature drop between the source and gate ensembles from data (circles) and numerical calculations (solid line).  Error bars indicate the $95\%$ confidence interval for the temperature fits. d) Calculated chemical potential drop using data (circles) and numerical calculations (solid line) from (a--c).  The vertical dashed line indicates the threshold feedback parameter, $\upsilon_\mathrm{BEC}$, above which a BEC forms in the gate well. Error bars indicate the error propagated from data in (b) and (c).}
\end{center}
\end{figure}

To study the transport dynamics between the source and gate wells, we observe the system using in-trap absorption images at a fixed evolution time of $\Delta t = 30~\mathrm{ms}$, at which point the gate has reached a quasi-steady-state. Holding the gate-source barrier height fixed at $V_\mathrm{GS} = 30~\mathrm{kHz}$, the gate-drain barrier height is varied. In order to extract quantitative data from in-trap absorption images, we strongly saturate the probe transition ($I_\mathrm{p} \gg I_{sat}$)~\cite{Reinaudi:2007}.  In the saturated probe regime we gain access to the source and gate populations, but lose information regarding the atoms in the drain.  The results of this experiment are shown in Fig.~\ref{fig:gate_turnon}. The measured total gate well atom number, $N_\mathrm{g,sat}$, is plotted in Fig.~\ref{fig:gate_turnon}(a) as a function of the feedback parameter, $\upsilon \equiv (V_\mathrm{GD}-V_\mathrm{GS})/kT_\mathrm{s}$. The feedback parameter describes the relative height of the two barriers, which controls the fraction of the source-gate current that is reflected back towards the source by the gate-drain barrier.  Strong probe saturation reduces the measured atom number in the gate well, yet the data shows the rapid increase in the population of the gate well beginning at the feedback parameter $\upsilon \approx -0.25$.  We calculate the temperature of the atoms within the source and gate wells by fitting a Gaussian to the thermal tails of the in-trap radial distribution. The temperature drop, $\tau \equiv (T_\mathrm{s}-T_\mathrm{g})/T_\mathrm{s}$, characterizes the temperature gradient between the source and gate wells, and is plotted in Fig.~\ref{fig:gate_turnon}(c).  The temperature drop data shows that $\tau > 0$ for $\upsilon > 0$, indicating that atoms in the gate are actually colder than those in the source.  

\section{\label{sec:Intro}Analysis and discussion\protect\\}
To model the transport dynamics in the triple-well system, we numerically calculate the steady-state ensembles in the source and gate wells.  Following previous works that model similar transport processes~\cite{Caliga:2015,Roos:2003tp,Walraven:1996qd}, we take the atom current flowing from the $i-$th to $j-$th well to be of the form:
\begin{equation}
\label{eqn:current}
I_{ij} = \gamma_i N_{\mathrm{th},i}\mathrm{exp}\left[-(V_{ij}-V_{0,i}-\mu_i)/kT_i\right].
\end{equation}
Here, the presence of the chemical potential reflects its role as a bias, which can be used to control current flow in much the same way, in fact, that an applied voltage can be used to control thermionic emission in electronics~\cite{Sze:1969sd,Bethe:1942}. The chemical potential associated with the $i-$th well has a contribution from its ground state energy as well as a thermodynamic contribution.  The ground state portion is explicitly accounted for by the parameter $V_{0,i}$, which is determined by the potential landscape.  The thermodynamic portion, $\mu_i$, is positive with the presence of a condensate and negative otherwise. When a BEC is present, $\mu_i$ is calculated based on the TF approximation for a condensate in a harmonic trap in the case of the gate well and a half-harmonic trap for the source and drain wells. In the absence of a BEC, the chemical potential can be calculated according to self-consistent mean field methods~\cite{Dalfovo:1999}. However, in this model the minimum single particle energy is $V_{0,i}$ and thus negative $\mu_i$ are set to zero. Within Eq.~(\ref{eqn:current}), the role of both $\mu_i$ and $V_{0,i}$ is to shift the minimum energy of the thermal atom distribution in the  $i-$th well, relative to the barrier height, $V_{ij}$. Therefore, the current depends on the trapping potential, the thermodynamic variables of the well from which the current originates, and the collision rate, $\gamma_i$, of thermally excited atoms in the well.  For the temperature regime considered here, $T_\mathrm{c} \gtrsim T > T_\mathrm{0}$, where $T_\mathrm{0}$ is the temperature associated with the interaction energy per particle, the collision rate is given by $\gamma_i = 32\pi^2\zeta(3/2)m(a_s k T_i)^2/h^3$~\cite{Pethick:2002tn}.   Within the expression for the collision rate, $\zeta$ is the Riemann-zeta function, $m$ is the atomic mass, $a_s$ is the s-wave scattering length, and $h$ is the Planck constant. Due to the height and width of the barriers, effects due to tunneling can be neglected and are omitted from the model. From Eq.~(\ref{eqn:current}) the net current between wells is calculated by adding all currents into and out of a given well. For example, the net source-gate current is given by $I_{sg,\mathrm{net}} = I_{sg}-I_{gs}$. As zero current originates from the drain well due to the presence of the terminator beam, current conservation dictates that $I_{sg} = I_{gs}+I_{gd}$. Note that upper versus lower case subscripts for the source, gate, and drain wells are used to distinguish between properties of the trapping potential, which are fixed parameters, and those of the atomic ensembles in each well and the currents, which are dynamical quantities. Using the data shown in Fig.~\ref{fig:atom_number}(a), the current into the gate well is determined.  We compare the measured current, determined from the decrease in total source atom number, to the result of Eq.~(\ref{eqn:current}) calculated using the measured $\mu_\mathrm{s}$ and $T_\mathrm{s}$. After an initial transient period of $\sim 20~$ms that occurs after lowering the height of the gate-source barrier to $V_\mathrm{GS} = 30$ kHz, we find that Eq.~(\ref{eqn:current}) accurately describes the experimentally measured current (see Fig.~\ref{fig:atom_number}(b)). Atom currents flow only along the longitudinal direction of the triple-well potential. Therefore, the energy current is given by
\begin{equation}
\label{eqn:energy}
\dot{E}_{ij} = I_{ij}(V_{ij}+\kappa k T_{i}),
\end{equation}
where the factor $\kappa \approx 2.9$ describes the excess energy carried by each atom that contributes to the current~\cite{Roos:2003tp}.
Using Eqs.~(\ref{eqn:current}) and~(\ref{eqn:energy}), and enforcing both particle and energy conservation, we numerically determine the steady-state source and gate well atom numbers and temperatures.  Repeating this process for various values of $\upsilon$, we reconstruct the gate atom number as a function of the feedback parameter.  The results are shown by the solid line in Fig.~\ref{fig:gate_turnon}(b) and exhibit qualitative agreement with the trend of the measured gate well population in Fig.~\ref{fig:gate_turnon}(a). The temperature drop is also extracted from the numerical calculation and is shown in Fig.~\ref{fig:gate_turnon}(c), and overlaid with the experimental data.  

Using the measured gate atom number and temperature, we determine the gate chemical potential.  Data in Fig.~\ref{fig:gate_turnon} was acquired using strongly saturated absorption imaging where $I_\mathrm{p} \gg I_{sat}$ and the effective scattering cross section is intensity dependent.  To first order, the correction to the OD  when strongly saturating the probe transition is a dimensionless coefficient~\cite{Reinaudi:2007}.  Assuming such a scale factor, $C$, between the measured gate well atom number, $N_\mathrm{g,sat}$, and the numerically calculated steady-state ($N_\mathrm{g} = C N_\mathrm{g,sat}$) we determine $C = 9.3(1)$. The scaled gate atom number is shown as open circles in Fig.~\ref{fig:gate_turnon}(b).  Deviation between the model prediction and scaled experimental data at large feedback parameters is a result of the increasing atom density in the gate. Large densities lessen the reduction of the scattering cross section because the cloud is optically thicker. As a result, the experimental data for the gate atom number at large positive feedback parameters is less than the numerically calculated population. The inset of Fig.~\ref{fig:gate_turnon}(b) shows the initial growth dynamics of the scaled atom number as a function of the evolution time, for $\upsilon = 0.6$.  The scaled gate atom number reaches the numerical value for the steady-state population by $\sim20$ ms.  In conjunction with the agreement between the measured source-gate current and Eq.~(\ref{eqn:current}) after $\sim20$ ms, equilibration of the gate well population verifies our assumption that the system is in quasi-steady-state by $\Delta t = 30~$ms. Using the scaled atom number from Fig.~\ref{fig:gate_turnon}(b) and temperature data from Fig.~\ref{fig:gate_turnon}(c), the chemical potential in the gate well is calculated in the TF limit.  The chemical potential drop, $\hat{\mu}_\mathrm{sg} \equiv (\mu_\mathrm{s}-\mu_\mathrm{g})/kT_\mathrm{s}$, characterizes the chemical potential gradient between the source and gate wells, and is shown in Fig.~\ref{fig:gate_turnon}(d).  For values of $\upsilon > 0.2$ the chemical potential drop is negative, revealing that the chemical potential in the gate well exceeds that of the source well. Furthermore, we determine the critical feedback parameter, $\upsilon_\mathrm{BEC} = -0.24$, above which a condensate forms in the gate well in steady-state. Below this threshold ($\upsilon < \upsilon_\mathrm{BEC}$) the chemical potential drop reflects the positive chemical potential of the source well relative to $V_\mathrm{0,G}$.

It is worth elaborating further on the presence of a negative chemical potential gradient between the source and gate wells. At first glance it may appear that in order to sustain a positive net source-gate current in this operating regime, work must constantly be done on the system. However, current flows due to gradients in both chemical potential and temperature. Thus, although the chemical potential gradient is negative, there is a positive temperature gradient that sustains current flow from source to gate. Additionally, a negative gradient in the chemical potential represents the conversion of thermal energy in the source to chemical potential energy in the gate. This process decreases the local entropy of the system; however, the presence of the terminator beam makes this an open system such that globally the entropy of the system increases and the second law of thermodynamics is upheld.

The data in Fig.~\ref{fig:gate_turnon} show that by tuning the feedback parameter it is possible to achieve quasi-steady-state ensembles in the source and gate that are related by temperature and chemical potential drops.  Drawing analogy to electronic transistor operation, the resulting chemical potential drop demonstrates self-biasing behavior controlled via the parameters of the trapping potential.  Electronic transistor functionality largely depends on the bias or quiescent operating point set by the voltages at each of its three terminals. We demonstrate both positive and negative chemical potential drops between the source and gate wells indicating operation in forward and reverse-bias modes, respectively.  Control over the temperature and chemical potential drops represents one method to manipulate the operation of an atomtronic transistor realized in a triple-well potential.   

\section{\label{sec:Intro}Conclusion\protect\\}
We have studied the transport of ultracold atoms at finite temperature in a triple-well potential from the perspective of atom currents flowing in an atomtronic device.  Dynamics of the atom current are directly observed in-trap using our atom chip based system in conjunction with a high numerical aperture imaging system.  Thermal and chemical potential gradients across a pair of repulsive barriers are shown to drive the flow of atoms in an effort to bring the total system into equilibrium.  However, we preclude global equilibrium by superimposing a terminator beam on the drain well to introduce a local source of dissipation. Removal of atoms that enter the drain simulates the condition in which the drain is impedance matched to a subsequent atomtronic circuit component.  By extinguishing the terminator beam prior to imaging the atoms in-trap, we gain access to the longitudinal spatial distribution of the gate-drain current. The spatial distribution corresponds to the energy spectrum of the current, which is on the order of the temperature of the source and gate ensembles.  While beyond the scope of this paper, one could characterize the power supplied by the triple-well system.

We also measured the quasi-steady-state population in the gate well and showed it to be controlled by the feedback parameter, which describes the relative height difference of the static barriers that separate the three wells. The initial growth of the gate well population is attributed to interatomic interactions that redistribute the energy of atoms as they traverse the gate well such that atoms become trapped and eventually thermalize.  Measured thermodynamic variables of the source and gate well were compared to numerically calculated steady-state values. From these results we showed that not only the temperature, but also the chemical potential of the steady-state gate ensemble is controlled via the feedback parameter. Despite being fed by an atom current that possesses energies in excess of the height of the gate-source barrier, positive values of the feedback parameter result in a gate well ensemble that is colder and has higher chemical potential than the source well ensemble.  Thus, we have demonstrated the ability to control the source-gate temperature and chemical potential drops in analogy to applying a voltage bias to the source-gate junction in an electronic transistor.  

\ack
This work was supported by the Air Force Office of Scientific Research (FA9550-14-1-0327), the National Science Foundation (PHY1125844), and by the Charles Stark Draper Laboratories (SC001-0000000759).


\begin{thebibliography}{13}
\bibitem{Nom:Trans} In labeling the regions of the triple-well potential, field-effect transistor nomencleture is used as it describes the role of each well in an atomtronic transistor: Atoms flow from the source, through a gate region, and into a drain well where they are removed from the system.  This naming scheme is meant to convey an intuition for the role of each region, rather than imply the behavior of the triple-well system is more analogous to the behavior of a field-effect transistor than a bipolar junction transistor.
\end{thebibliography}
\end{document}